\documentclass[twocolumn,showpacs,preprintnumbers,amsmath,amssymb]{revtex4}

\usepackage{graphicx}

\usepackage{dcolumn}

\usepackage{bm}

\usepackage{color}

\usepackage{xspace}

\newcommand{\etal}{\latin{et~al}.\@\xspace}
\usepackage[usenames,dvipsnames]{xcolor}

\newcommand{\latin}[1]{\emph{#1}}

\usepackage{epstopdf}

\newcommand{\beq}{\begin{equation}}
\newcommand{\eeq}{\end{equation}}
\newcommand{\beqn}{\begin{eqnarray}}
\newcommand{\eeqn}{\end{eqnarray}}

\newcommand{\fig}{Fig.\ }

\begin{document}

\title{Percolation mechanism drives actin gels to the critically connected state}
\author{Chiu Fan Lee}
\email{c.lee@imperial.ac.uk}
\address{Department of Bioengineering, Imperial College London, South Kensington Campus, London SW7 2AZ, United Kingdom}
\author{Gunnar Pruessner}
\email{g.pruessner@imperial.ac.uk}
\affiliation{Department of Mathematics, Imperial College London,  South Kensington Campus, London SW7 2AZ, United Kingdom}
\date{\today}
\begin{abstract}
Cell motility and tissue morphogenesis depend crucially on the dynamic remodelling  of actomyosin networks. An actomyosin network consists of an actin polymer network connected by crosslinker proteins and motor protein myosins that generate internal stresses on the network.
A recent discovery shows that for a range of experimental parameters, actomyosin networks contract to clusters with a power-law size distribution [Alvarado J. {\it et al.} (2013) Nature Physics {\bf 9} 591]. Here, we argue that actomyosin networks can exhibit robust critical signature without fine-tuning because the dynamics of the system can be mapped onto a modified version of percolation with trapping (PT), which is known to show critical behaviour belonging to the static percolation universality class without the need of fine-tuning of a control parameter. We further employ our PT model to generate experimentally 
testable predictions.
\end{abstract}
 \maketitle 

\section{Introduction}
Actomyosin networks constitute an archetypal example of active matter \cite{julicher_physrep07,joanny_hfspj09}. 
In a typical experimental setup, an actomyosin network consists of 
self-assembled actin filaments of variable length are connected {\it via} crosslinkers and myosins \cite{koenderink_pnas09}. 
{ Actin filaments are two-stranded helical polymers of the protein
actin, with a diameter of around 6 nm \cite{alberts_b08}. The protein
fascin binds to two actin filaments and serves as a crosslinker to give
connectivity to the whole network at high enough concentration
\cite{koenderink_pnas09}. The molecular motor myosin II, generates internal stresses in the network by converting ATP to mechanical forces \cite{alberts_b08}. In biological systems, myosins play a dominate role in muscle contraction and are responsible for the motility of eukaryotic cells.
}
An actomyosin network can undergo contraction due to the internal stress generated by the myosins pulling two oppositely oriented actin filaments past each other \cite{alberts_b08,lenz_prx14}.  Such an active system has been shown to exhibit diverse patterns \cite{kruse_prl04,voituriez_prl06}. Besides its intrinsic scientific interest, studying  actomyosin networks is essential to our understanding of cell motility, and cell and tissue morphogenesis \cite{clarke_annrev77,munja_dev14}. 
Recently, it was found that a planar actin network can be remodelled by myosins in a way that exhibits robust scale invariant structures similar to what is observed in static percolation at the critical point \cite{alvarado_natphys13}.
One of the most intriguing findings is that the characteristics of critical phenomena observed do not  require fine-tuning of any control parameter, which suggests the presence of {\it bona fide} self-organised criticality \cite{bak_prl76,pruessner_b12,watkins_r15}. In addition to the experimental findings, the authors of \cite{alvarado_natphys13} propose a 
detailed theoretical model that 
successfully accounts for the salient features of their experimental observations. Their model takes into consideration the various microscopic interactions among actin filaments, myosins and crosslinkers. 

In the present work, we asked a different question: what is the minimal physical model that explains the apparent self-organised critical nature of the active actin network observed?
This question is strongly motivated by the expectations that at criticality, many microscopic details of the model can become irrelevant \cite{wilson_sciam79,cardy_b96}.
Below, we will argue that the underlying mechanism behind the observed critical behaviour of actomyosin network corresponds to a kind of percolation process known as percolation with trapping (PT) \cite{dias_jpa86}, rather than self-organised criticality as envisaged by Bak, Tang, and Wiesenfeld \cite{bak_prl76,pruessner_b12,watkins_r15}.
Our model not only recovers the most relevant experimental features, but
allows in addition to relate them to established results in the theory
of percolation.

The structure of the paper is as follows:
In Sect.~\ref{experimental_result}, we recapitulate the essential experimental findings in \cite{alvarado_natphys13}.
In Sect.~\ref{model} we will argue that the underlying biophysical mechanism can be mapped onto a variant of percolation with trapping. In Sect.~\ref{discussion}
we provide numerical evidence that the model indeed belongs to the static percolation universality class.
We then derive predictions from our model, which include how the cluster size depends on the time of production, and the correlation between cluster sizes and their distances from the boundary.

\section{Experimental Findings}\label{experimental_result}
\begin{figure}
\begin{center}
\includegraphics[scale=.5]{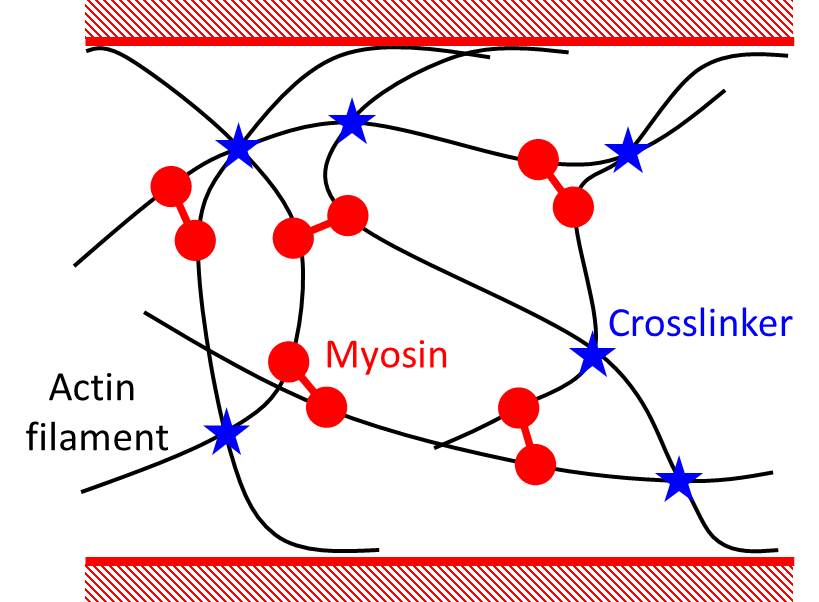}
\end{center}
\caption{\label{experiment_cartoon}
Schematics of the actin meshwork studied in the experiment by Alvarado \etal \cite{alvarado_natphys13} (see also their Fig.~1(a)). Some actin filaments (black lines) connect to the top and to the bottom boundary. Interconnects are provided by crosslinking fascins (blue stars). Molecular motors myosins (red dumbbells) pull the filament together. The  experiment was on a quasi-two dimensional plane of size $2.5\times2.5$mm$^2$, and  was observed over the course of about two hours.
}
\end{figure}

In their experiments, Alvarado {\it et al.} \cite{alvarado_natphys13} prepared a quasi-two dimensional network of actin filaments, about $2.5\times2.5$mm$^2$ in size, which are connected by fascin crosslinkers, illustrated schematically in Fig.~\ref{experiment_cartoon}. Myosin motors pull the filaments together, thereby exerting contractile forces on the network \cite{lenz_prx14}. The resulting  crosslinker unbinding and actin filament movement fundamentally alter the underlying actin network  (see supplementary movies in \cite{alvarado_natphys13}).

By tracing the trajectories of the actin filaments in the course of the contraction, the authors were able to reconstruct the connectivity of the actin network throughout the dynamical process. Specifically, depending on the number of crosslinkers in the system $M_c$, three different regimes are observed \cite{alvarado_natphys13}:
\begin{enumerate}
\item
At low $M_c$, upon activation of the myosins, the actin network is 
contracted into many foci. By retracing the position of the actin molecules back to $t=0$, one observes that the resulting foci originated from actin clusters with areas of similar sizes. 
\item
At intermediate $M_c$, the retracing to the original areas indicates a power law distribution in sizes with an exponent of $-1.91$, which is similar to expected exponent of $-187/91$ in random percolation at criticality \cite{stauffer_b94}.
\item
At high $M_c$, the actin network contracts to one single piece with the boundary of the experimental container discernible.
\end{enumerate}
The fact that the intermediate regime, in which scale invariant size distribution is found, exists for a wide range of crosslinker concentration suggests that the critical signature occurs without fine-tuning, as observed in systems displaying self-organised criticality.

\begin{figure}
\begin{center}
\includegraphics[scale=.6]{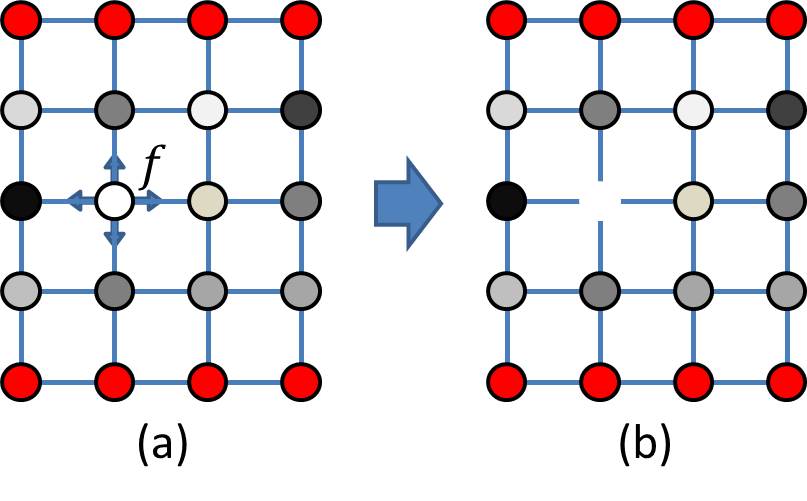}
\end{center}
\caption{(a) The nodes adhered to the top and bottom boundaries are red and with adhesion strength $s_A$. The interior nodes (grey) have connection strengths drawn from  a Poisson distribution and the higher the strength, the darker the colour of the node. Each node is supposed to have a force $f$ pulling on it on all sides and the weakest node will disappear first (b). 
}
\label{fig:1}
\end{figure}

\section{Model}\label{model}
Alvarado \etal \cite{alvarado_natphys13} developed a model that incorporates
the microscopic details of the actin filaments under load, and the binding and unbinding kinetics of the crosslinkers. In the present work, we aim to account for the scale invariant nature of the experimental observations by using a minimal model. The motivation behind this task comes from the general expectation that at criticality, many microscopic details become irrelevant. To obtain such a minimal description, we first list our key assumptions and our interpretations of the particular roles of the various  ingredients in the experimental system.

\subsection{Initialisation of the actin network}
\begin{itemize}
\item
We assume that the actin filaments form a planar network. For simplicity, we assume the network formed is a regular square lattice with $N=L\times L$  nodes.
\item
The addition of crosslinkers serve to connect the filaments at the nodes in a random fashion. To model the randomness involved, we assign a strength value $s_k$ to  node $k$ that is chosen from a Poisson distribution with mean $\bar{s}\sim M_c/N$. This is to imitate the random number of crosslinkers connected at each node.
\item
Contractile forces generated by the molecular motor myosins induce pulling force $f_k$ on node $k$. Similar to the above, we assume that at the initial time, the pulling force is Poisson distributed with mean $\bar{f}\sim M_m/N$ where $M_m$ is the number of myosin motors in the system. 
\item
Besides the pulling forces within the system, there is the adhesion that connects the actin network to the boundary. We assume here that the boundary adhesion strength is $s_A$ and following the modelling method in  \cite{alvarado_natphys13}, we assume that the actin networks adhere to the top and bottom boundaries only. Note that although  the adhesion strength here is in principle random, the exact values do not affect our analysis.
\end{itemize}
The initial setup of the system is shown schematically in \fig \ref{fig:1}(a). We now describe the mysosin induced dynamics on the actin network.

\begin{figure}
\begin{center}
\includegraphics[scale=.5]{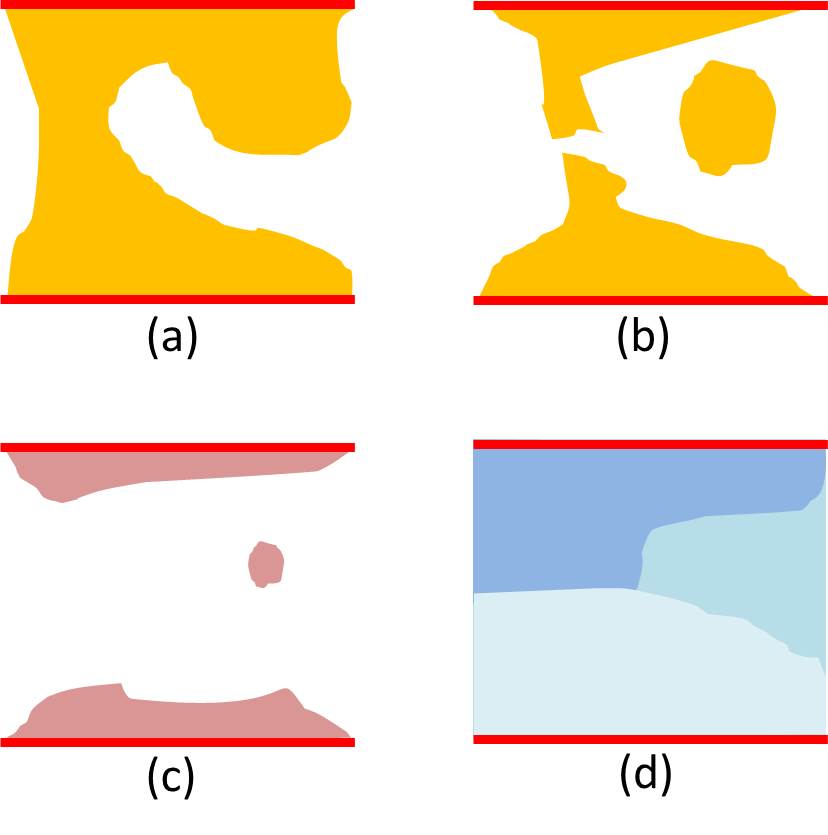}
\end{center}
\caption{(a) As random deletion of nodes start, the network starts to shrink. (b) As soon as a connected cluster detached from either top or bottom boundaries, it quickly shrinks to a focal point or to the boundary, and as a result, node deletion is no longer possible (c). (d) Retracing the path of each  node, the areas at the initial time corresponding to the three disjoint, fragmented clusters can be obtained, which are schematically shown in (d).
}
\label{fig:2}
\end{figure}

\subsection{Dynamics}
\begin{itemize}
\item
As the mysoin pulling force starts to act, we assume that the nodes rip (and thus effectively disappear) sequentially according to the reverse order of their unbinding propensity. The disappearance propensity of node $k$ is given by a function $B(s_k,f_k)$ such that $B$ decreases with $s_k$ but increases with $f_k$. 
Physically, the disappearance of a node amounts to having the crosslinkers on that node unbind from the actin filaments at that node due to loading (\fig \ref{fig:1}(b)). After a node disappears, the pulling forces are modified to achieve the force balance in the system   \cite{alvarado_natphys13}. 
This implicitly assumes that force balance equilibration is fast compared to crosslinker unbinding \cite{alvarado_natphys13}. In other words, we assume that the unbinding times scale $\tau_u$ is much greater than the force balance equilibration time scale $\tau_f$: $\tau_u \gg \tau_f$.
\item
Besides the mysoin unbinding events, connected clusters of actin filaments can also contract, which we assume to occur whenever a connected cluster is detached from {\it either} boundary (\fig \ref{fig:2}). We also assume that the contraction time scale $\tau_c$ is small compared to the unbinding timescale $\tau_u$ ($\tau_u \gg \tau_c$), so that once a cluster is free to contract, it contracts to a focal point almost immediately before the next unbinding event occurs. Once detached from the boundary the cluster freely contracts and all forces due to mechanical tension cease.
The contracted cluster is thus removed from further dynamical evolution of the system, i.e., node deletions no longer occur in this cluster.
\end{itemize}

Based on the above model ingredients, we can now explain why the system is partitioned into three regimes as $M_c$ varies: if $M_c$ is small, many nodes will be empty and so the actin network would not be connected as whole, i.e.,
 the network does not percolate. As a result, the actin network fragment into many small foci. On the other hand, if $M_c$ is high, then the average strength of a node may be higher than the adhesion strength ($\bar{s} > s_A$). As a result, the whole network detaches from the adhesion boundary due to the internal contraction, which is observed experimentally. The interesting regime is the intermediate range of $M_c$, where the actin network percolates and the crosslinker unbinding occurs predominately in the interior. This is the regime we will focus on from now on.

Instead of incorporating all of the above elements in the modelling, we will make the crucial simplification here by eliminating the dynamical elements in the system that are of time scale shorter than the time scale of node deletion in the network.  In other words, we will ignore
the force equilibration and cluster collapse steps.  In addition, we assume that the pulling force experienced by  every node is identical at each time step.  
 In this mean-field approximation, we are effectively assuming that although the pulling force at each node may vary and fluctuate as rupture occurs throughout the mesh, the spatial correlation of the force fluctuations remains short ranged so that upon coarse graining, a well-defined mean value exists. 
As long as the (initial) forces do not exceed the adhesion strength $s_A$ to detach the mesh as a whole, and as long as they are strong enough to produce effectively random rupture throughout the system, the crosslinker unbinding events will  produce random bond deletion between cluster. 
The actual form of the unbinding propensity function $B$ is therefore irrelevant as far as static properties are concerned. Indeed, the resulting fragmentation can be simulated as follows:
\begin{quotation}
Start with a fully connected squared lattice and delete interior nodes at random. If a connected cluster detaches from either boundary, then it is taken out the system and its configuration recorded. 
 \end{quotation}
We will refer to this as the ``rupturing model''. We note that a model based on random deletion of bonds in the context of active gels has also been proposed recently \cite{sheinman_pre15}. 
 The main difference between our model and that in \cite{sheinman_pre15} is that our model incorporates the effect of boundary adhesion into the model description, which we believe is of experimental significance. In fact, we believe that both models are variants of the percolation with trapping model detailed below, where boundary conditions play a key role in the model dynamics.
 
\subsection{Relation to percolation with trapping}
The scheme above is a variant of (invasion) percolation with trapping \cite{dias_jpa86}. Invasion percolation (\emph{without} trapping) \cite{wilkinson_jpa83} generates a single cluster by successively occupying more and more sites from an initial seed on an infinite lattice, thought to be occupied by the invader. Assigning initially a random number drawn uniformly and independently from the unit interval, the next site to be invaded is the one with the smallest such random number among non-invaded sites neighbouring invaded sites. Islands of non-invaded sites surrounded by invaders may appear in the process but disappear again as the invasion progresses. 
 Invasion percolation \emph{with} trapping suppresses these events, as no further invasion is allowed once an ``island'' of non-invaded sites is fully enclosed by invaders.
  As a motivation, consider extracting oil in a porous medium by pumping water (invader) in one specific site \cite{dias_jpa86}. The oil content in a particular pocket can be extracted as long  as the pocket is connected to the extraction site. 
However, if the oil pocket is completely surrounded by water, the oil content will be locked inside and thus be insusceptible to further displacement by water.

 Although originally motivated by invasion percolation, occupation by invaders does not need to be restricted to sites neighbouring to invaded sites. When occupation by invaders can
 take place spontaneously throughout the lattice, except inside isolated islands of unoccupied sites, this is known as percolation with trapping (PT) \cite{dias_jpa86}.


The rupturing model as described above may be thought of as a version of PT in reverse. Nodes disappear in a random sequence,
but no further nodes are deleted inside a cluster once it is detached from a boundary.
The key difference between traditional PT and the rupturing model above is this:
While in traditional PT \emph{unoccupied} clusters stop shrinking as soon as they become detached from the boundaries, the rupturing (disappearance of occupied sites) stops as soon as an \emph{occupied} cluster becomes detached from a boundary. 
 In other words, our evolution algorithm is based on clusters of \emph{occupied} sites becoming disconnected from the boundaries and we analyse the statistics of \emph{occupied} sites, whereas the traditional approach is based on \emph{unoccupied} sites becoming disconnected from the boundaries, while still taking the statistics of \emph{occupied} sites.


A second difference, which numerically turns out to be insignificant, is that in our model, clusters are already taken out of the system if they detach from {\it any} of the adhesion boundaries, Fig.~\ref{fig:2}. This is to reflect the experimental reality, as it was observed that clusters attached to only one adhesion boundary will contract to a the boundary quickly \cite{alvarado_natphys13}.
Although many big clusters therefore contract to the boundaries, smaller ones form earlier in the process as they get cut off from the bulk.

 \begin{figure}
 \begin{center}
 \includegraphics[scale=.52]{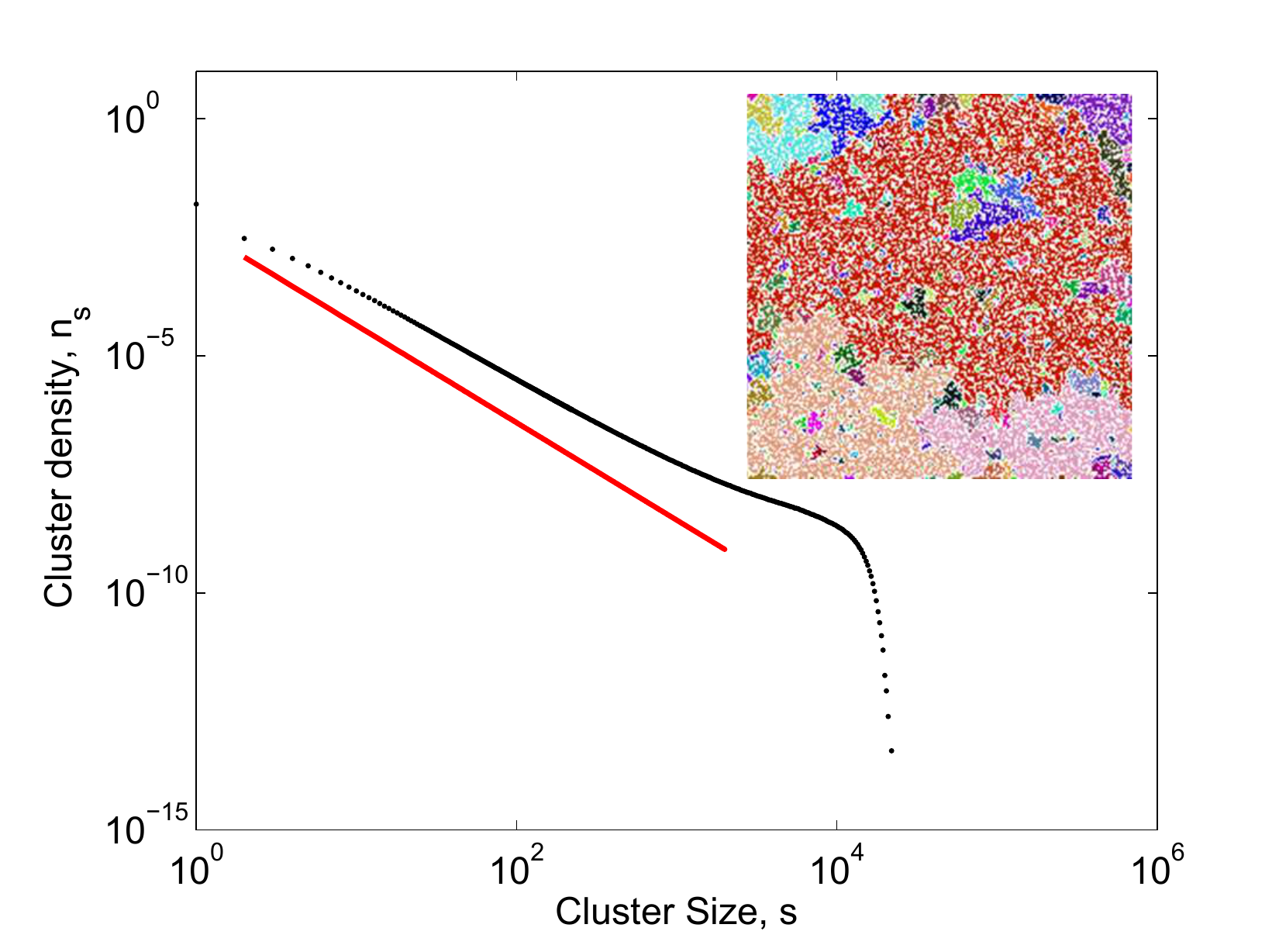}
 \end{center}
 \caption{{\it Cluster density distribution.} Simulation of our PT model on a 200$\times$200 square lattice produces a power-law distribution of cluster density ($n_s$) with an exponent close to 2 (black curve). The red curve corresponds to the power law distribution with exponent $-187/91 \simeq -2.05$ which corresponds to the exponent of static percolation on a 2D square lattice \cite{stauffer_b94}. The inset figure shows the clusters generated from our model.
 }
 \label{res1}
 \end{figure}

 \begin{figure}
 \begin{center}
 \includegraphics[scale=.52]{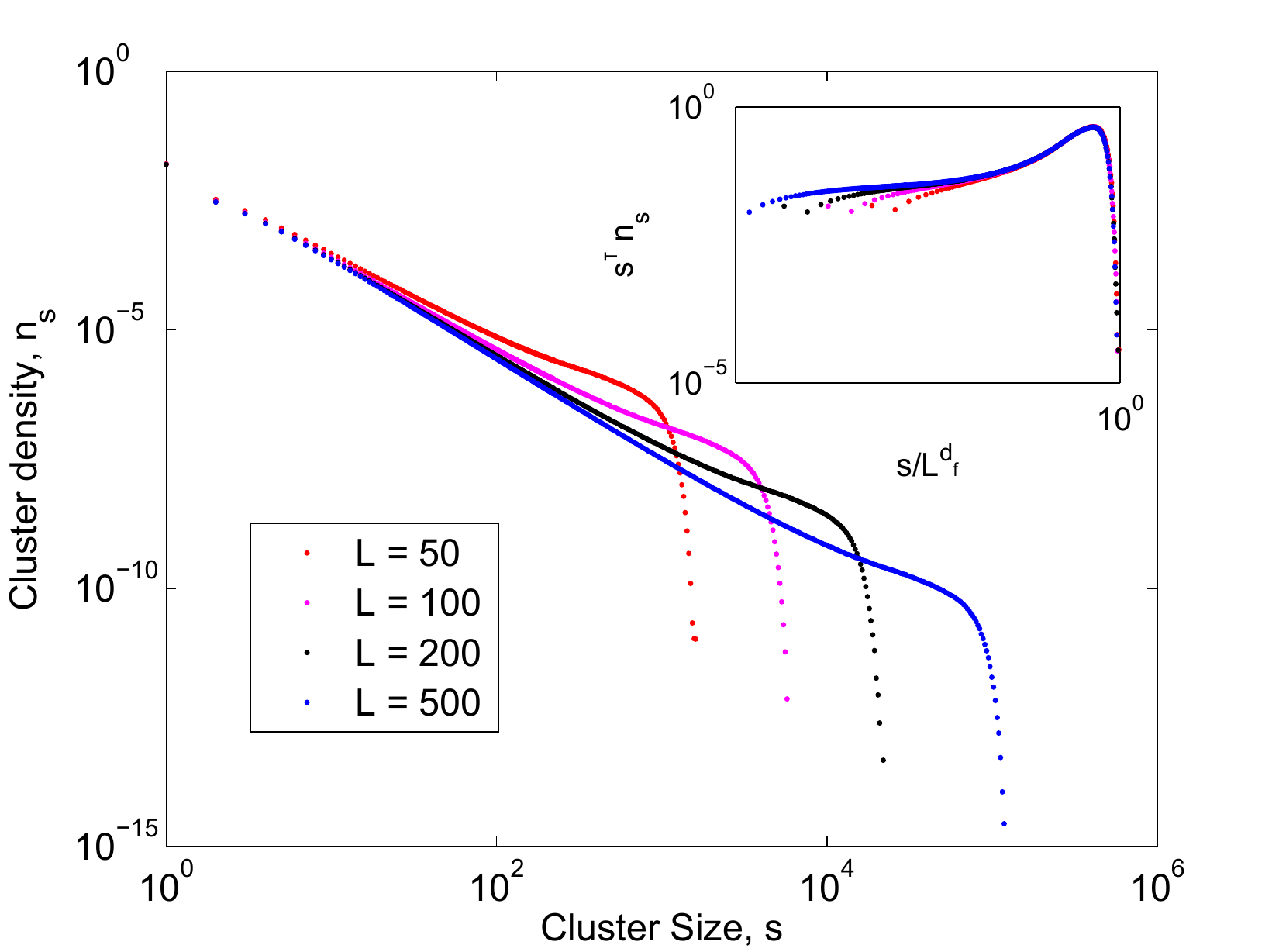}
 \end{center}
 \caption{{\it Finite size scaling.} Simulations of our PT model on different sizes of square lattice. The inset shows that cluster density distribution collapse onto a single curve as predicted by static percolation theory, which dictates that $s^\tau n_s = \Phi (s/L^{d_f})$
 where $\Phi$ is a scaling function and the exponents are: $\tau = 187/91$ and $d_f = 91/48$ \cite{stauffer_b94}.
 }
 \label{fig:res2}
 \end{figure}

\section{Results and discussion}\label{discussion}
In this Section we will first show that our rupturing model reproduces the scale invariant nature of the experimental observations. We will then present several experimentally verifiable predictions derived from our model simulations.

\subsection{Static percolation universality class}
\begin{figure}
\begin{center}
\includegraphics[scale=.53]{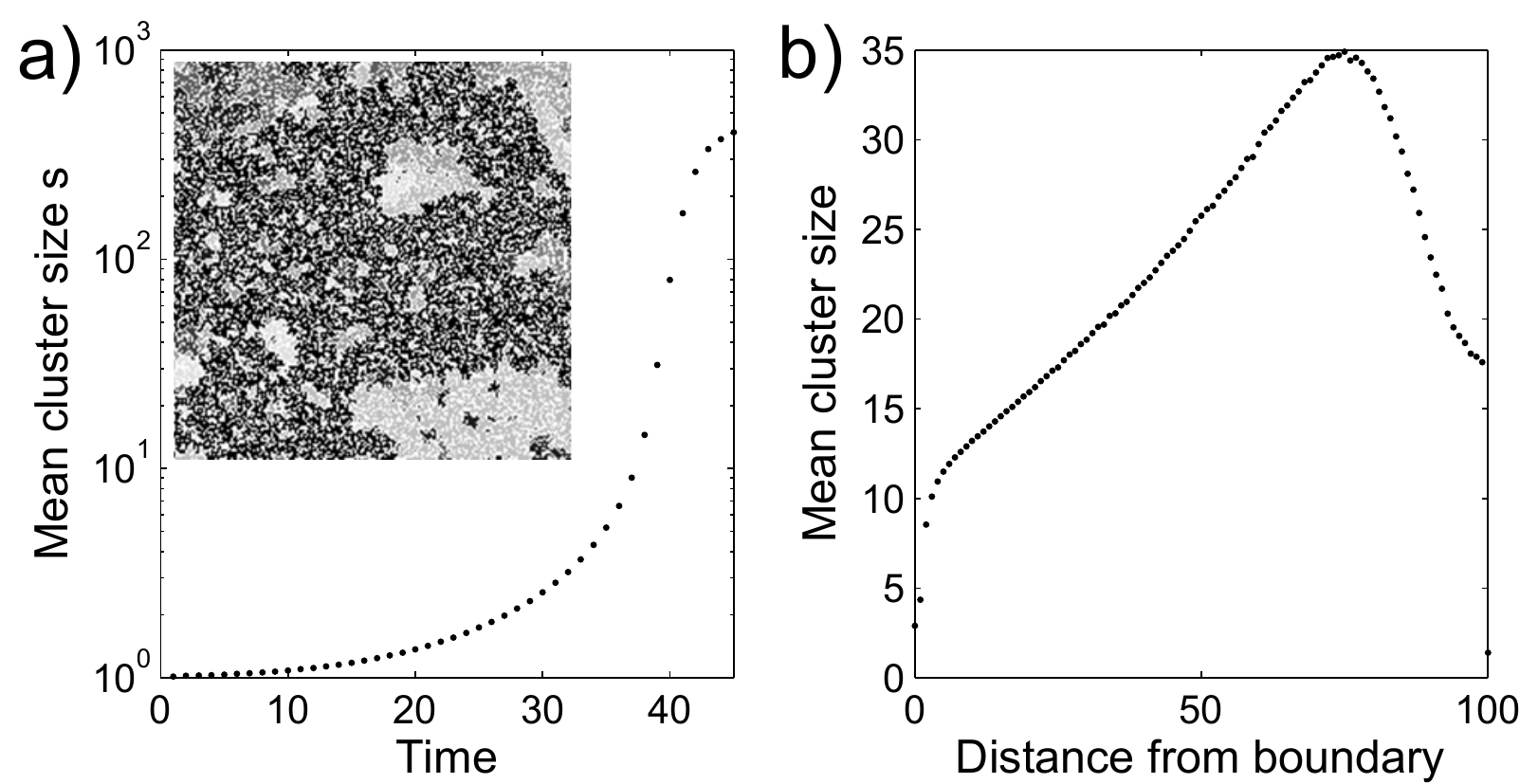}
\end{center}
\caption{{\it Predictions.} a) Size of clusters as a function of the time of their free contractions. The inset shows an example of the temporal ordering (light to dark) of cluster contractions. b) Size of clusters as a function of the distance between their centres of mass and the closest adhesion boundary. Results are from simulations performed on a $200 \times 200$ square lattice.
}
\label{fig:res3}
\end{figure}

Numerical simulations of the original PT model have amply supported that PT belongs to the static percolation universality class  \cite{dias_jpa86}. In terms of cluster size distribution, Fig.~\ref{res1} shows that our modified PT also preserves the  characteristics of static critical percolation on a 2D square lattice. 
As demonstrated in Fig.~\ref{fig:res2}, 
although the absolute value of the power law scaling obtained numerically seems to be smaller than the expected exponent of $187/91\simeq 2.05$, this is likely due to finite size effects. The collapse shown in the inset of Fig.~\ref{fig:res2} using exponents predicted by static percolation theory clearly demonstrates that the modified PT model presented here belongs to the universality class of static percolation \cite{stauffer_b94}. 
We have also studied lattices with aspect ratios other than unity (the
aspect ratio used in Figs.~\ref{res1}--\ref{fig:res3}) and found the same
scaling behaviour, as expected from universality.

We note that starting from a critically connected percolation network, a no-enclaves version of  the cluster size counting method has  recently been proposed
\cite{sheinman_prl15,sheinman_pre15}. Using this counting method, the
power law exponent governing the no-enclave cluster density as a
function of cluster size is found to be lower than $\tau \simeq 2.05$,
which is thus  closer to the experimentally measured exponent of $1.91$
\cite{alvarado_natphys13}. In relation to our model, since the
no-enclave modification is ultimately based on a percolation network at
criticality \cite{PruessnerLee:2016}, which is exactly the network generated by our model, the no-enclave modification can also be applied to our model straightforwardly.

\subsection{Predictions}

We have so far argued that for a range of experimental parameters, the contractile dynamics of an actomyosin network can be mapped onto a  modified version of PT, which enables us to explain i) why critical behaviour is observed generically without fine tuning of a control parameter, and ii) why the exponent observed in the  cluster size distribution is close to that predicted by static percolation theory \cite{alvarado_natphys13}. We will now use our PT model to generate predictions that await experimental verifications.

{\it Bigger clusters collapse later.} Although the dynamical evolution of our model is purely sequential and physical time scales do not feature, we expect that the temporal trend of how the sizes of cluster produced vary as the simulation progresses reflects what happens experimentally. As such, our first prediction is that within the critical domain (i.e, at the moderate range of crosslinker concentration $M_c$), larger and larger clusters collapse onto foci as time progresses (\fig \ref{fig:res3}(a)).
This prediction can be intuitively understood by the following argument: If large clusters collapse first, the network is unlikely to remain connected between the two adhesion boundaries, which would imply the end of the dynamical evolution since no nodes would be connected to both boundaries and no further rupture occurs.

{
{\it Cluster center of mass distribution.} \fig \ref{fig:res3}(b) shows
that on average, the centers of mass of larger clusters are further away
from the nearest boundary, which can be trivially explained by the
larger physical size of the clusters. A more interesting prediction here
is that there is a peak in the curve shown  \fig \ref{fig:res3}(b),
which indicates that the centers of mass for large clusters stay away
from the mid-line between the two adhesion boundaries.  One plausible
explanation is that large clusters are asymmetric in shape in such a way
that a larger proportion of their nodes are closer to the adhesive
boundary that they remain attached to. This asymmetry is built into the
model by construction as a cluster is formed when it is disconnected
from one adhesion boundary while still being connected to the other. In other words, the cluster
``rips at the thin end'', which implies that there is a thinned and a
bulkier side.  This predicted cluster shape asymmetry
remains to be verified experimentally.  
}

\section{Summary \& Outlook}
Contracting actomyosin networks can exhibit robust power-law distribution of cluster sizes \cite{alvarado_natphys13}. The formation of the clusters is due to the tearing apart of the initially homogeneous actin filament network by the contractile forces generated by the molecular motor myosins. Here, we provide a minimal model to account for the robustness of the power-law scaling observed experimentally. Specifically, we argue that the key ingredients in the  microscopic dynamics of  an actomyosin network can be accounted for by  a modified version of percolation with trapping. We show that this rupturing model exhibits critical behaviour that belongs to the static percolation universality class without fine-tuning. Furthermore, we generated specific predictions based on our PT model, which await experimental verifications.

As far as the suggested connection to self-organised criticality is concerned, our model displays a form of self-organisation, as further rupturing stops when the clusters detach from the boundaries, i.e., close to the percolation threshold. This happens provided time scales are sufficiently separated. The scale invariant features are those found in static percolation. With a widely accepted definition of self-organised criticality in mind (such as non-trivial scale invariance, spatio-temporal power law correlation, self-tuning \cite{watkins_r15}) one might conclude that this really \emph{is} a case of genuine self-organised criticality. However, there is no intermittency, no stationarity and strictly no dynamics. There is, in fact, no ongoing process that would enable self-tuning by some feedback loop, because scale invariance occurs as further evolution terminates. Our model therefore shares with invasion percolation reservations that have been raised about it being an instance of self-organised criticality \cite{grassberger_jphysfrance90,sornette_jphysfrance95}.

In terms of future work, actomyosin networks constitute an archetype of active matter, which has been traditionally modelled as a continuous medium with internal stresses generated within each volume element \cite{toner_annphys05,marchetti_rmp13}. The fact that the medium on which the stresses are transmitted can be transformed by the internal stresses, as investigated here, challenges this traditional approach. Indeed, a theory of active matter on networks (rather than just lattices) would be required to properly account for the dynamical and mechanical properties of actomyosin networks in their full generalities. Another interesting future direction is to study the properties of myosins and crosslinkers  as the network evolves. Here, we focused only on the cluster size distribution resulting from the active contraction and found that we recovered the static percolation universality class. Active motility of particles in a continuous medium is known to lead to novel universal behaviour \cite{chen_njp15}. 
The question therefore arises whether the dynamics of myosins and crosslinkers produces novel dynamical critical behaviour.

\begin{acknowledgements}
The authors thank Jos\'{e} Alvarado, Gijsje Koenderink, Michael Sheinman, Abhinav Sharma and Fred MacKintosh for stimulating discussions.
\end{acknowledgements}


\end{document}